# On-chip superconducting GHz RF reflectometry of the capacitance response in bilayer graphene

Sung Jin An,[1,†] Minseo Cho,[2,†] Minjun Park,[2†] Dohun Kim,[2,†] HyeonJeong An,[1,2] Seung-Bo Shim,[3] Hakseong Kim,[3] Sunghun Lee,[4] Myoung-Jae Lee,[1] Kenji Watanabe,[5] Takashi Taniguchi,[6] Jungpil Seo,[2] Myunglae Jo,[7,*] Youngwook Kim,[2,*] Minkyung Jung[1,8,*]

[1] DGIST Research Institute, DGIST, Daegu 42988, Republic of Korea

[2] Department of Physics and Chemistry, DGIST, Daegu 42988, Republic of Korea

[3] Quantum Technology Institute, Korea Research Institute of Standards and Science, 34113 Daejeon, Republic of Korea

[4] Industrial AX Innovation Research Institute and Sensor Technology Support Center, DGIST, Daegu 42988, Korea

[5] Research Center for Materials Nanoarchitectonics, National Institute for Materials Science, 1-1 Namiki, Tsukuba 305-0044, Japan

[6] Research Center for Electronic and Optical Materials, National Institute for Materials Science, 1-1 Namiki, Tsukuba 305-0044, Japan

[7] Department of Physics, Kyungpook National University, Daegu 41566, Republic of Korea

[8] Department of Interdisciplinary Engineering, DGIST, Daegu 42988, Republic of Korea

[†]These authors contributed equally.

[*]Corresponding authors

Email: myunglae.jo@knu.ac.kr, y.kim@dgist.ac.kr, minkyung.jung@dgist.ac.kr

**ABSTRACT**

In dual-gated bilayer graphene, a perpendicular displacement field opens a band gap that modifies both the channel conductance and the electronic compressibility, motivating measurements that resolve resistive and capacitive responses on the same device. We integrate an hBN-encapsulated bilayer graphene heterostructure with an on-chip superconducting Nb lumped-element $LC$ resonator and carry out RF reflectometry near 4.25 GHz. DC transport and finite-bias spectroscopy on the same device provide a transport reference. Top and bottom gates independently set the carrier density and displacement field. The DC and RF gate maps share the same gate-dependent features, with finite-bias measurements revealing a region of suppressed conductance whose bias extent grows with the displacement field, consistent with a field-induced gap. The gate-dependent resonance-frequency shift is converted to the effective capacitance seen by the resonator using an equivalent-circuit model. The capacitance shows a minimum near the conductance-suppressed region, consistent with reduced electronic compressibility in the gapped bilayer graphene, and exhibits an electron-hole asymmetry. The on-chip configuration probes the gate-dependent admittance of a dual-gated van der Waals heterostructure, providing capacitance-sensitive information that complements DC transport within a single device.

## INTRODUCTION

Bernal-stacked bilayer graphene (BLG) is a gapless van der Waals semimetal whose low-energy band structure is controlled by a perpendicular displacement field [1–4]. A finite displacement field breaks inversion symmetry between the two graphene layers and opens a band gap at the charge neutrality point [5–10], with the gap size set by the interlayer potential difference and reaching above 100 meV in graphite-gated, hBN-encapsulated devices [10]. This combination of electrostatic band-gap control and low disorder has made BLG an important system for low-dimensional materials physics and for gate-defined quantum devices, including quantum dots [11–19], quantum point contacts [13,20], and spin- or valley-based qubit architectures [15,18].

Characterizing gate-defined BLG devices and the field-induced gapped regime calls for measurements that go beyond DC transport. The field-induced gap modifies the density of states and electronic compressibility, both of which shift the device capacitance in addition to the dissipative conductance. High-frequency readout also extends the bandwidth available for tracking charge configurations in BLG quantum devices beyond standard low-frequency lock-in techniques. Microwave-based methods, including superconducting cavities [21–23] and lumped-element *LC* matching networks [24–28], address this regime by embedding the complex admittance of the device into a resonant circuit; RF reflectometry then translates gate-dependent modulations of the channel's resistance and capacitance into variations in the reflected microwave signal. With sensitivity to both reactive and dissipative impedance, reflectometry has become an established technique for fast charge sensing, shot-noise acquisition, and dispersive readout in nanoscale architectures [27–29].

In BLG devices, RF reflectometry has so far been implemented mainly with MHz-range lumped-element matching circuits, including gate-defined quantum dots and dual-gated devices using microscale graphite gates on insulating substrates [24–26]. These studies established that the reflected RF signal tracks gate-dependent changes in the BLG conductance, device capacitance, or charge configuration. However, MHz tank circuits based on external or normal-metal circuit elements are limited by resistive loss and parasitic capacitance arising from wiring, bonding pads, and device packaging [27], which restrict the accessible resonance frequency and complicate impedance matching in high-resistance 2D devices. Superconducting microwave resonators and cavities have recently been coupled to graphene and BLG devices to probe charge dynamics, capacitance, and compressibility at microwave frequencies [21–23]. Compared with extended cavities, a lumped-element *LC* resonator provides a compact on-chip geometry that can be placed close to the BLG channel and incorporated into a dual-gate device layout, and on-chip superconducting *LC* resonators have been used for GHz-frequency impedance matching of high-resistance mesoscopic devices [29]. What remains less explored is the use of an on-chip superconducting lumped-element resonator to characterize a dual-gated BLG heterostructure near the displacement-field-induced gapped regime, where the conductance and capacitance both depend on the displacement field.

To this end, we couple a graphite-gated BLG channel to a superconducting Nb resonator operating in the gigahertz range. Microwave reflection through the source contact reports on the channel admittance, which carries both the conductance and capacitance of BLG. The reflected signal is interpreted within a lumped-circuit framework, providing a single-chip route to follow the BLG load through the gapped regime.

## RESULTS

### Resonator-BLG device and measurement setup

Figure 1(a) shows our device, in which the dual-gated BLG heterostructure is assembled on top of a pre-patterned Nb *LC* resonator and shares its source line with the resonator's matching node. The BLG channel then enters the resonator as a gate-tunable load, and its admittance variations are read out as changes in the reflected microwave amplitude $|S_{11}|^2$ [29,30]. Figure 1(b) recalls the band-structure schematic, with the interlayer potential difference $\Delta$ controlled by the perpendicular displacement field. This gate-controlled gap sets the regime probed by the DC, RF, and capacitance measurements described below.

The superconducting resonator is fabricated from a 100-nm-thick Nb film sputtered on a high-resistivity Si substrate (> 10 k$\Omega$·cm), chosen to reduce microwave loss at GHz frequencies. Electron-beam lithography and $SF_6$/Ar reactive-ion etching define a planar spiral inductor, whose distributed capacitance to the surrounding metal and substrate forms the parasitic capacitance $C_S$ of the lumped-element *LC* resonator. The multi-turn spiral has micrometer-scale linewidth and spacing, and is designed to place the resonance near 4.25 GHz. The inset of Fig. 1(a) shows an SEM image of a representative resonator. The central metal island acts as the bonding pad for the microwave line, while the outer end of the spiral is routed to the BLG source contact so that the channel admittance loads the resonant circuit.

Assembly of the BLG heterostructure follows the van der Waals dry-transfer and AFM cleaning method [31-33]. A Bernal-stacked bilayer graphene flake is encapsulated in bottom- and middle-hBN, with a graphite flake placed above the middle-hBN as the top gate and a further top-hBN capping layer, using a polymer-based dry-transfer pick-up process. The completed stack is transferred onto the pre-patterned resonator chip. One-dimensional edge contacts are then formed by electron-beam lithography, $CF_4/O_2$ reactive-ion etching to expose the graphene edge, and Cr/Au deposition with a thickness of 10/120 nm. In the final device, the superconducting *LC* resonator is connected directly to the source electrode of the BLG channel. From the two-terminal resistance of the BLG channel in the conducting regime, the series resistance is estimated to be on the order of 1 kΩ. This estimate provides the scale of contact and access contributions to the measured dissipative response.

Figure 1(c) shows the cross-section of the device. From bottom to top, the stack consists of an embedded Ti/Au bottom gate in $SiO_2$, a bottom hBN dielectric (b-hBN), the BLG channel, a middle hBN dielectric (m-hBN), a graphite top gate, and a top hBN capping layer (t-hBN). This planar architecture keeps the heterostructure nearly flat and reduces topographic steps that could affect device uniformity. A graphite flake incorporated during the stacking process serves as the top gate ($V_T$), while the embedded bottom electrode provides the bottom gate ($V_B$), enabling independent electrostatic control of the channel. The carrier density and displacement field follow $n = (c_{TG}V_T + c_{BG}V_B)/e$ and $D = (c_{TG}V_T - c_{BG}V_B) / (2\varepsilon_0)$, where $c_{TG}$ and $c_{BG}$ are the per-unit-area capacitances of the top and bottom gates and $\varepsilon_0$ is the vacuum permittivity. Unless otherwise noted, the measurements were performed at $T = 2$ K, well below the superconducting transition of Nb ($T_c \approx 9.2$ K) and with a thermal energy $k_BT \approx 170$ μeV that is small compared to the displacement-field-induced gap.

The RF and DC wiring used for the reflectometry measurements is shown in Fig. 1(a). A microwave tone from a vector network analyzer (VNA) is sent to the sample through cryogenic attenuators and coupled to the superconducting resonator through a directional coupler. The same coupler separates the reflected microwave signal, which is amplified at room temperature and recorded as $|S_{11}|^2$. A bias tee connected to the source electrode combines the microwave excitation with the DC bias line, so that RF reflectometry and transport measurements are performed in the same device configuration. The RF port of the bias tee is connected to the *LC* resonator, while the DC port applies the source-drain bias voltage $V_{SD}$. The drain electrode is held at virtual ground.

**Equivalent-circuit model and dual-gate RF response**

We analyze the RF response of this device using an equivalent-circuit framework. The model, shown in Fig. 2(a), describes the superconducting resonator connected to the BLG device. The resonator branch consists of an inductance $L$ in series with an effective loss resistance $R_{loss}$, with a parasitic capacitance $C_S$ shunting the inductor node to ground; $C_S$ arises from the distributed capacitance of the spiral inductor and the surrounding substrate, while $R_{loss}$ captures the internal loss (residual conductor, dielectric, and radiative contributions). On the load side, the BLG channel contributes a dissipative admittance through the channel resistance $R_G$ and a capacitive admittance through the total device capacitance $C_t$. $C_t$ itself contains both the geometric gate capacitance and the finite electronic compressibility of BLG.

In a series-capacitance picture,

$$C_t^{-1} = C_G^{-1} + C_Q^{-1} \text{ with } C_Q = e^2 A\rho(E_F), \quad (1)$$

Here $C_G$ is the effective geometric capacitance of the dual-gate stack. For the present device, both gates are DC-biased through bias tees and act as AC grounds at the resonator frequency, so $C_G$ is approximated as $C_G \approx A \times (c_{TG} + c_{BG})$. $\rho(E_F)$ is the density of states per unit area at the Fermi energy [34-37].

To include dielectric loss in the capacitive branch, we introduce a shunt conductance $G_S = \omega \cdot C_T \cdot \tan(\delta)$ in parallel with $C_T$, where $\omega = 2\pi f$ and $\tan(\delta)$ is the dielectric loss tangent. This term describes dissipation associated with the capacitive gate path and is distinct from $R_G$, which represents dissipative transport through the BLG channel. For the high-quality hBN dielectric used here, $\tan(\delta)$ is small at gigahertz frequencies, so $G_S \ll \omega C_T$ and the capacitive branch is dominated by $C_T$.

The total input impedance of the circuit is then written as

$$Z_{in} = j\omega L + R_{loss} + \left[\frac{1}{R_G} + j\omega C_T + \omega C_T tan\delta_{eff}\right]^{-1} \quad (2)$$

$$\text{where, } C_T = C_t + C_S$$

In the low-loss limit, the resonance frequency is approximated by

$$f_r = \frac{1}{2\pi\sqrt{LC_T}} \qquad (3)$$

with the full impedance evaluated otherwise.

A change in $C_T$ then shifts the resonance frequency, whereas a change in $R_G$ mainly modifies the damping, depth, and linewidth of the resonance. The microwave reflection coefficient is

$$\Gamma \equiv \frac{Z_{in} - Z_0}{Z_{in} + Z_0} \qquad (4)$$

where $Z_0$ is the characteristic impedance of the transmission line, and the measured reflected power is $|S_{11}|^2 = |\Gamma|^2$. The reflected signal therefore provides a gate-dependent probe of the BLG load admittance, without separating resistance and quantum capacitance in the RF response.

Figure 2(b) displays the measured $|S_{11}|^2$ spectra at three representative gate voltages, ($V_T$, $V_B$) = (2.0 V, -1.41 V), (2.0 V, -1.32 V), and (2.0 V, -1.05 V), corresponding to $G$ = 0.036, 0.350, and 0.416 $e^2/h$, respectively. The resonance dips at $f_0 \approx 4.251$ GHz with a loaded quality factor $Q_L \approx 27.3$ (FWHM $\Delta f \approx 154$ MHz). Across these three gate configurations the resonance shifts by $\Delta f_r \approx 8.75$ MHz, corresponding to a fractional shift $\Delta f_r/f_0 \approx 2.1 \times 10^{-3}$. Both the resonance frequency and the dip magnitude shift as the gates tune the channel, consistent with the equivalent-circuit model above: the frequency shift tracks the effective capacitance $C_T$, while the dip depth and linewidth track gate-dependent dissipation in $R_G$.

The circuit parameters are determined by fitting the measured reflection spectra to the calculated input impedance $Z_{in}$ and the corresponding reflection coefficient, Eqs. (2) and (4) (see Supporting Information for details). In the high-conductance regime, the model reproduces the measured spectra well and gives $L$ = 18.49 nH, $\tan\delta_{eff}$ = 0.0085, and $R_{loss}$ = 29.5 Ω, with an effective capacitance $C_T$ of approximately 76 fF.

After establishing the resonator response, we compare DC transport and RF reflectometry over the dual-gate parameter space. Figure 2(c) plots the DC current ($I$) measured as a function of $V_T$ and $V_B$ at a fixed source-drain bias $V_{SD}$ = 1 mV. A diagonal trace of suppressed current marks the charge-neutrality condition of the dual-gated BLG channel, broadening for gate configurations with larger perpendicular displacement field, in agreement with the opening of a field-induced gap in BLG [5–10]. Away from this trace, the current increases as the Fermi level is tuned into the electron- or hole-conducting bands.

The RF response, recorded at a fixed frequency near $f_0 \approx 4.251$ GHz (indicated in Fig. 2(b)) and at an incident power of −70 dBm, is plotted as the reflected power $|S_{11}|^2$ in Fig. 2(d). The map reproduces the gate-dependent structure of the DC current, including the diagonal feature at charge neutrality and the broadened response at finite displacement field, so that the BLG admittance loads the superconducting resonator and the fixed-frequency reflectometry tracks changes in the device admittance across the dual-gate map.

**Finite-bias spectroscopy of the gapped regime**

To examine whether the suppressed region in the dual-gate maps reflects a field-induced gapped state, we turn to finite-bias spectroscopy. The DC current and the RF reflected signal are recorded as functions of $V_{SD}$ and $V_B$ at three fixed top-gate voltages, $V_T = -1.2$, $-1.4$, and $-1.6$ V, covering different displacement-field conditions near charge neutrality. Figure 3 shows the DC current maps in panels (a), (c), and (e), alongside the simultaneously measured reflected power $|S_{11}|^2$ in panels (b), (d), and (f).

In the DC maps, a region of suppressed transport appears around zero source-drain bias, consistent with the field-induced gap in dual-gated BLG [5–10]. The bias extent of this region widens as the displacement-field condition is increased. The simultaneously measured RF maps reproduce the corresponding gate- and bias-dependent structure, showing that the microwave response tracks changes in the BLG load admittance in the finite-bias regime. A weaker zero-bias dip remains visible outside the main suppressed region and likely arises from contact-related resistance. We use the finite-bias data as evidence for the gapped regime and as a reference for interpreting the RF capacitance response below, rather than to extract a precise band-gap value.

**Gate-dependent capacitance response**

Figure 4(a) presents the differential conductance, $G$, as a function of $V_B$ and $V_{SD}$ at $V_T = 2$ V, chosen to place the channel well inside the gapped regime. A deep suppression of conductance appears around $V_B = -1.4$ to $-1.5$ V, marking the gate-voltage range where the Fermi level is tuned through the gapped region. To capture the corresponding microwave response, we measure the reflected signal $|S_{11}|^2$ as a function of frequency and $V_B$ along the line marked A in Fig. 4(a), at fixed $V_T = 2$ V and $V_{SD} = 5$ mV. The finite bias was chosen to track the resonance shift along a stable line across the conductance-suppressed region; we therefore interpret the extracted $C_T$ as a bias-conditioned capacitance response of the device. In Fig. 4(b), the resonance dip shifts in frequency and changes in amplitude near the conductance-suppressed region, in line with the model expectation that $C_T$ shifts the resonance while $R_G$ modifies the dip depth and linewidth. The conductance line cut in Fig. 4(c), taken along the same path, provides the transport reference for this gate-voltage range.

Using the extracted inductance ($L$ = 18.49 nH) and the gate-dependent resonance frequency $f_r(V_B)$, we obtain the effective capacitance $C_T$ from $C_T(V_B) = 1/[(2\pi f(V_B))^2 L]$. Here $f_r(V_B)$ is determined by fitting each reflection spectrum to a standard one-port resonator model, with the remaining circuit parameters fixed at the values established in the high-conductance regime (Supporting Information). Because this procedure relies primarily on the resonance-frequency shift, it is less sensitive to imperfect modeling of the full line shape in the low-conductance regime, where dissipative effects, localized states, and charge inhomogeneity can affect the resonance depth and linewidth. The resulting $C_T(V_B)$, with uncertainties propagated from the standard error of the fitted $f_r$, is shown in Fig. 4(d).

The extracted capacitance is the effective capacitance $C_T = C_S + C_t$, where $C_S$ is the gate-independent parasitic capacitance and $C_t$ is the BLG device capacitance defined in Eq. (1). $C_T$ develops a clear minimum near the conductance minimum. Because $C_S$ is gate-independent, this modulation arises entirely from $C_t$, the series combination of the geometric capacitance $C_G \approx A(c_{TG} + c_{BG}) \approx 52.7$ fF and the quantum capacitance $C_Q$. Since the observed modulation is small compared with $C_G$, the reduction near the gap is consistent with a suppressed quantum-capacitance contribution (equivalently, reduced electronic compressibility) rather than a change in $C_G$.

The minimum remains finite, which may reflect residual compressibility from disorder, localized gap states, or spatial charge inhomogeneity [34-38]. The trace is also asymmetric between electron and hole branches, possibly reflecting a near-layer effect in layer-polarized BLG. In this picture, the band-edge states are expected to acquire different weights on the two graphene sheets for opposite carrier polarity under a perpendicular displacement field [34–38], and the resulting asymmetry in their electrostatic coupling to the gates and in the interlayer screening leaves an imprint on the measured response. Other sources of asymmetry, such as contact-related asymmetry or residual disorder, may also contribute, so we regard the near-layer capacitance effect as a possible interpretation rather than a unique assignment. The resonance-frequency shift thus yields information on the device response in the gapped BLG regime, complementing the DC conductance.

## CONCLUSION

Low-frequency current maps localized the displacement-field-induced gap, and the microwave reflection followed the same gate dependence, indicating that the reflected signal responds to the gate-dependent admittance of the BLG channel, including dissipative and capacitive contributions. From the resonance-frequency dependence we obtained the device capacitance as a function of gate voltage; the capacitance dip near the transport gap is consistent with a suppression of the quantum-capacitance contribution, with asymmetric behavior between the electron- and hole-side carriers that may reflect the layer-polarized band-edge picture. The measurement therefore brings dispersive and dissipative information together on a single chip. This circuit-on-chip approach can be extended to other graphene-based and carbon-based van der Waals heterostructures, where the small-signal admittance provides information complementary to conventional transport.

## References

[1] E. McCann, M. Koshino, The electronic properties of bilayer graphene. Rep. Prog. Phys. **76**, 056503 (2013).

[2] E. McCann, V. I. Fal'ko, Landau-level degeneracy and quantum Hall effect in a graphite bilayer. Phys. Rev. Lett. **96**, 086805 (2006).

[3] F. Guinea, A. H. Castro Neto, N. M. R. Peres, Electronic states and Landau levels in graphene stacks. Phys. Rev. B **73**, 245426 (2006).

[4] K. S. Novoselov, E. McCann, S. V. Morozov, V. I. Fal'ko, M. I. Katsnelson, U. Zeitler, D. Jiang, F. Schedin, A. K. Geim, Unconventional quantum Hall effect and Berry's phase of 2π in bilayer graphene. Nat. Phys. **2**, 177–180 (2006).

[5] E. V. Castro, K. S. Novoselov, S. V. Morozov, N. M. R. Peres, J. M. B. Lopes dos Santos, J. Nilsson, F. Guinea, A. K. Geim, A. H. Castro Neto, Biased bilayer graphene: Semiconductor with a gap tunable by the electric field effect. Phys. Rev. Lett. **99**, 216802 (2007).

[6] J. B. Oostinga, H. B. Heersche, X. Liu, A. F. Morpurgo, L. M. K. Vandersypen, Gate-induced insulating state in bilayer graphene devices. Nat. Mater. **7**, 151–157 (2008).

[7] H. Min, B. Sahu, S. K. Banerjee, A. H. MacDonald, Ab initio theory of gate induced gaps in graphene bilayers. Phys. Rev. B **75**, 155115 (2007).

[8] Y. Zhang, T. T. Tang, C. Girit, Z. Hao, M. C. Martin, A. Zettl, M. F. Crommie, Y. R. Shen, F. Wang, Direct observation of a widely tunable bandgap in bilayer graphene. Nature **459**, 820–823 (2009).

[9] E. Icking, L. Banszerus, F. Wörtche, F. Volmer, P. Schmidt, C. Steiner, S. Engels, J. Hesselmann, M. Goldsche, K. Watanabe, T. Taniguchi, C. Volk, B. Beschoten, C. Stampfer, Transport spectroscopy of ultraclean tunable band gaps in bilayer graphene. Adv. Electron. Mater. **8**, 2200510 (2022).

[10] E. Icking, D. Emmerich, K. Watanabe, T. Taniguchi, B. Beschoten, M. C. Lemme, J. Knoch, C. Stampfer, Ultrasteep slope cryogenic FETs based on bilayer graphene. Nano Lett. **24**, 11454–11461 (2024).

[11] M. T. Allen, J. Martin, A. Yacoby, Gate-defined quantum confinement in suspended bilayer graphene. Nat. Commun. **3**, 934 (2012).

[12] A. M. Goossens, S. C. M. Driessen, T. A. Baart, K. Watanabe, T. Taniguchi, L. M. K. Vandersypen, Gate-defined confinement in bilayer graphene-hexagonal boron nitride hybrid devices. Nano Lett. **12**, 4656–4660 (2012).

[13] A. Kurzmann, H. Overweg, M. Eich, A. Pally, P. Rickhaus, R. Pisoni, Y. Lee, K. Watanabe, T. Taniguchi, T. Ihn, K. Ensslin, Charge detection in gate-defined bilayer graphene quantum dots. Nano Lett. **19**, 5216–5221 (2019).

[14] R. Garreis, C. Tong, J. Terle, M. J. Ruckriegel, J. D. Gerber, L. M. Gächter, K. Watanabe, T. Taniguchi, T. Ihn, K. Ensslin, W. W. Huang, Long-lived valley states in bilayer graphene quantum dots. Nat. Phys. **20**, 428–434 (2024).

[15] M. Eich, F. Herman, R. Pisoni, H. Overweg, A. Kurzmann, Y. Lee, P. Rickhaus, K. Watanabe, T. Taniguchi, M. Sigrist, T. Ihn, K. Ensslin, Spin and valley states in gate-defined bilayer graphene quantum dots. Phys. Rev. X **8**, 031023 (2018).

[16] L. Banszerus, S. Möller, E. Icking, K. Watanabe, T. Taniguchi, C. Volk, C. Stampfer, Single-electron double quantum dots in bilayer graphene. Nano Lett. **20**, 2005–2011 (2020).

[17] L. Banszerus, B. Frohn, A. Epping, D. Neumaier, K. Watanabe, T. Taniguchi, C. Stampfer, Gate-defined electron-hole double dots in bilayer graphene. Nano Lett. **18**, 4785–4790 (2018).

[18] A. O. Denisov, V. Reckova, S. Cances, M. J. Ruckriegel, M. Masseroni, C. Adam, C. Tong, J. D. Gerber, W. W. Huang, K. Watanabe, T. Taniguchi, T. Ihn, K. Ensslin, H. Duprez, Spin-valley protected Kramers pair in bilayer graphene. Nat. Nanotechnol. **20**, 494–499 (2025).

[19] L. Banszerus, S. Möller, C. Steiner, E. Icking, S. Trellenkamp, F. Lentz, K. Watanabe, T. Taniguchi, C. Volk, C. Stampfer, Spin-valley coupling in single-electron bilayer graphene quantum dots. Nat. Commun. **12**, 5250 (2021).

[20] H. Overweg, H. Eggimann, X. Chen, S. Slizovskiy, M. Eich, R. Pisoni, Y. Lee, P. Rickhaus, K. Watanabe, T. Taniguchi, V. Fal'ko, T. Ihn, K. Ensslin, Electrostatically induced quantum point contacts in bilayer graphene. Nano Lett. **18**, 553–559 (2018).

[21] M. J. Ruckriegel, L. M. Gächter, D. Kealhofer, M. Bahrami Panah, C. Tong, C. Adam, M. Masseroni, H. Duprez, R. Garreis, K. Watanabe, T. Taniguchi, A. Wallraff, T. Ihn, K. Ensslin, W. W. Huang, Electric dipole coupling of a bilayer graphene quantum dot to a high-impedance microwave resonator. Nano Lett. **24**, 7508–7514 (2024).

[22] K. Maji, J. Sarkar, S. Mandal, S. H., M. Hingankar, A. Mukherjee, S. Samal, A. Bhattacharjee, M. P. Patankar, K. Watanabe, T. Taniguchi, M. M. Deshmukh, Superconducting cavity-based sensing of band gaps in 2D materials. Nano Lett. **24**, 4369–4375 (2024).

[23] V. Ranjan, S. Zihlmann, P. Makk, K. Watanabe, T. Taniguchi, C. Schönenberger, Contactless microwave characterization of encapsulated graphene p-n junctions. Phys. Rev. Appl. **7**, 054015 (2017).

[24] L. Banszerus, S. Möller, E. Icking, C. Steiner, D. Neumaier, M. Otto, K. Watanabe, T. Taniguchi, C. Volk, C. Stampfer, Dispersive sensing of charge states in a bilayer graphene quantum dot. Appl. Phys. Lett. **118**, 093104 (2021).

[25] T. Johmen, M. Shinozaki, Y. Fujiwara, T. Aizawa, T. Otsuka, Radio-frequency reflectometry in bilayer graphene devices utilizing microscale graphite back-gates. Phys. Rev. Appl. **20**, 014035 (2023).

[26] M. Shinozaki, T. Johmen, A. Hosaka, T. Seo, S. Yashima, A. Shirachi, K. Noro, S. Sato, T. Kumasaka, T. Yoshida, T. Otsuka, RFSoC-based radio-frequency reflectometry in gate-defined bilayer graphene quantum devices. Appl. Phys. Express **18**, 075001 (2025).

[27] F. Vigneau, F. Fedele, A. Chatterjee, D. Reilly, F. Kuemmeth, M. F. Gonzalez-Zalba, E. Laird, N. Ares, Probing quantum devices with radio-frequency reflectometry. Appl. Phys. Rev. **10**, 021305 (2023).

[28] F. K. Malinowski, L. Han, D. de Jong, J. Wang, C. G. Prosko, G. Badawy, S. Gazibegovic, Y. Liu, P. Krogstrup, E. P. A. M. Bakkers, L. P. Kouwenhoven, J. V. Koski, Radio-frequency C-V measurements with subattofarad sensitivity. Phys. Rev. Appl. **18**, 024032 (2022).

[29] M. C. Harabula, T. Hasler, G. Fülöp, M. Jung, V. Ranjan, C. Schönenberger, Measuring a quantum dot with an impedance-matching on-chip superconducting *LC* resonator at gigahertz frequencies. Phys. Rev. Appl. **8**, 054006 (2017).

[30] S. J. An, J. Kim, M.-C. Jung, K. Park, J. Park, S.-B. Shim, H. Kim, Z. B. Siu, M. B. A. Jalil, C. Schönenberger, N. Myoung, J. Seo, M. Jung, Radio-frequency detection of Fabry-Pérot interference and quantum capacitance in long-channel three-dimensional Dirac semimetal $Cd_3As_2$ nanowires. ACS Appl. Electron. Mater. **7**, 9195–9203 (2025).

[31] L. Wang, I. Meric, P. Y. Huang, Q. Gao, Y. Gao, H. Tran, T. Taniguchi, K. Watanabe, L. M. Campos, D. A. Muller, J. Guo, P. Kim, J. Hone, K. L. Shepard, C. R. Dean, One-dimensional electrical contact to a two-dimensional material. Science **342**, 614-617 (2013).

[32] Y. Kim, P. Herlinger, T. Taniguchi, K. Watanabe, and J.H. Smet, "Reliable Postprocessing Improvement of van der Waals Heterostructures," ACS Nano 13, 14182–14190 (2019)

[33] D. Kim, S. Kim, Y. Cho, J. Lee, K. Watanabe, T. Taniguchi, M. Jung, J. Falson, Y. Kim, Full-dry flipping transfer method for van der Waals heterostructure. Curr. Appl. Phys. **59**, 165–168 (2024).

[34] E. A. Henriksen, J. P. Eisenstein, Measurement of the electronic compressibility of bilayer graphene. Phys. Rev. B **82**, 041412(R) (2010).

[35] A. F. Young, L. S. Levitov, Capacitance of graphene bilayer as a probe of layer-specific properties. Phys. Rev. B **84**, 085441 (2011).

[36] A. F. Young, C. R. Dean, I. Meric, S. Sorgenfrei, H. Ren, K. Watanabe, T. Taniguchi, J. Hone, K. L. Shepard, P. Kim, Electronic compressibility of layer-polarized bilayer graphene. Phys. Rev. B **85**, 235458 (2012).

[37] K. Kanayama, K. Nagashio, Gap state analysis in electric-field-induced band gap for bilayer graphene. Sci. Rep. **5**, 15789 (2015).

[38] Y. Kim, D. S. Lee, S. Jung, V. Skákalová, T. Taniguchi, K. Watanabe, J. S. Kim, J. H. Smet, Fractional Quantum Hall States in Bilayer Graphene Probed by Transconductance Fluctuations. Nano Lett. **15**, 7445-7451 (2015).

## Acknowledgements

This work is supported by the Mid-Career Researcher Program of NRF (RS-2023-NR076585 and RS-2023-00278511) and the DGIST R&D Program of the Ministry of Science, ICT, and Future Planning (26-ET-02, 26-SR-02). This work was supported by NRF (RS-2023-00269616, RS-2023-00258732, RS-2024-00402302, RS-2025-25463923, RS-2023-00285353, RS-2025-02317602, RS-2025-02315685, RS-2025-16070349, RS-2025-00519187, RS-2025-25444462, RS-2025-25464247) grants funded by the Korean government (MSIT & MOE). Y.K acknowledges support from National Research Foundation of Korea (NRF) (Grant No. RS-2025-00557717) and the Nano and Material Technology Development Program through the National Research Foundation of Korea (NRF) funded by Ministry of Science and ICT (No. RS-2024-00444725). We also acknowledge the partner group program of the Max Planck Society. Part of this work was supported by Global Partnership Program of Leading Universities in Quantum Science and Technology (RS-2025-02317602). K.W. and T.T. acknowledge support from the JSPS KAKENHI (Grant Numbers 21H05233 and 23H02052), the CREST (JPMJCR24A5), JST and World Premier International Research Center Initiative (WPI), MEXT, Japan.

## Author Contributions

S. J. A., M. C., M. P., and D. K. contributed equally. S. J. A., Y. K., and M. J. conceived the project. S. J. A, M. C., M. P., D. K., and M. C. carried out the device fabrication. The low-temperature measurements were done by S. J. A. and M. C. under supervision of M. J. Datasets were analyzed by S. J. A., M. J., Y. K., and M. J. T. T. and K. W. synthesized the h-BN crystals. All authors contributed to the manuscript writing.

## Competing Interests

The authors declare no competing interests

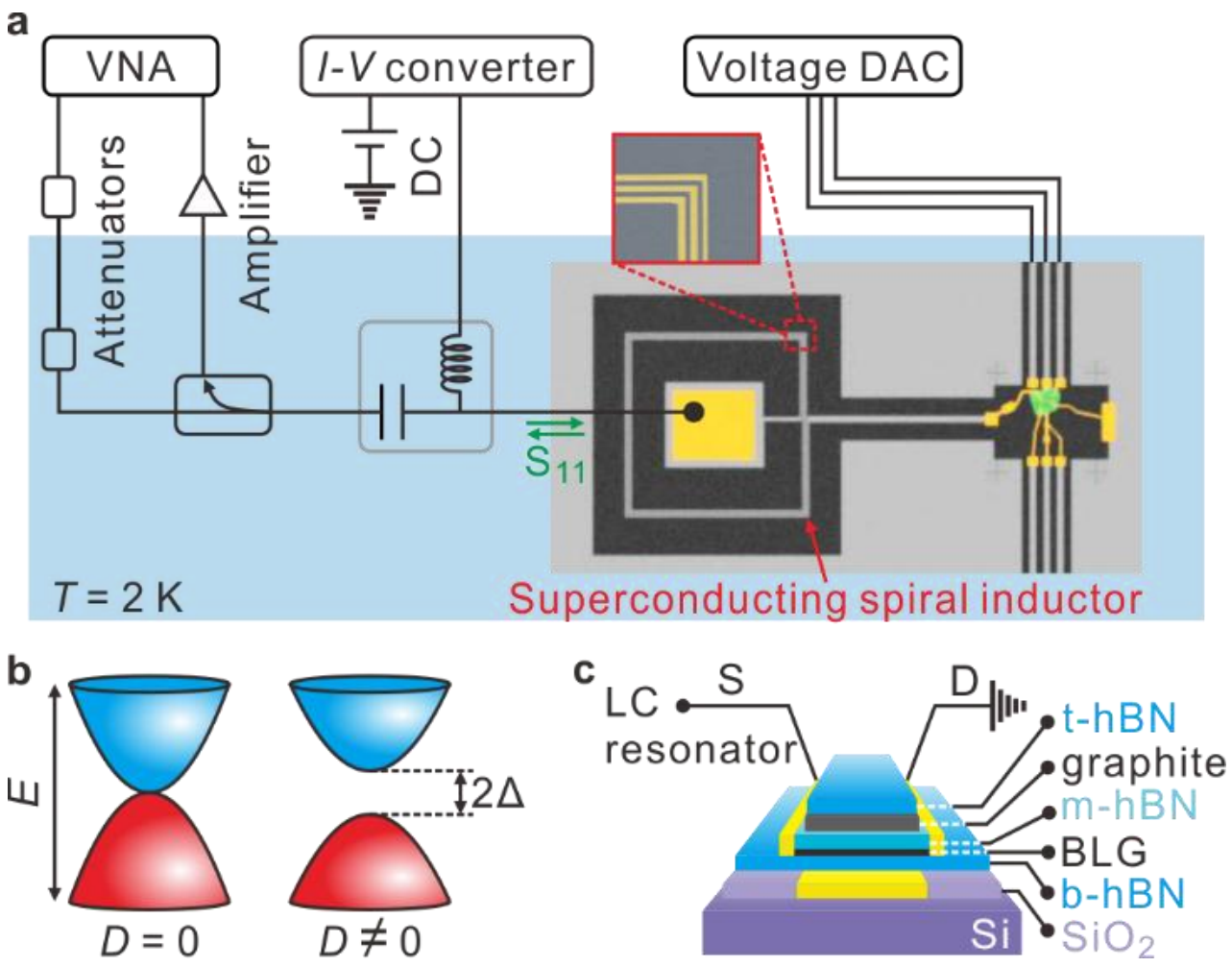

**Figure 1| Device architecture and RF measurement scheme.** (a) RF reflectometry setup for the BLG device connected to an on-chip superconducting Nb *LC* resonator. A bias tee combines the DC source-drain bias and microwave excitation at the source side, and the reflected signal is measured through a directional coupler. Inset: SEM image of a representative resonator-BLG device. (b) Schematic band structure of Bernal-stacked BLG without and with a perpendicular displacement field, showing the field-induced gap opening at the charge neutrality point. (c) Cross-section of the dual-gated BLG heterostructure: top hBN (t-hBN), graphite top gate, middle hBN dielectric (m-hBN), BLG channel, bottom hBN dielectric (b-hBN), and embedded Ti/Au bottom gate, with Cr/Au edge contacts.

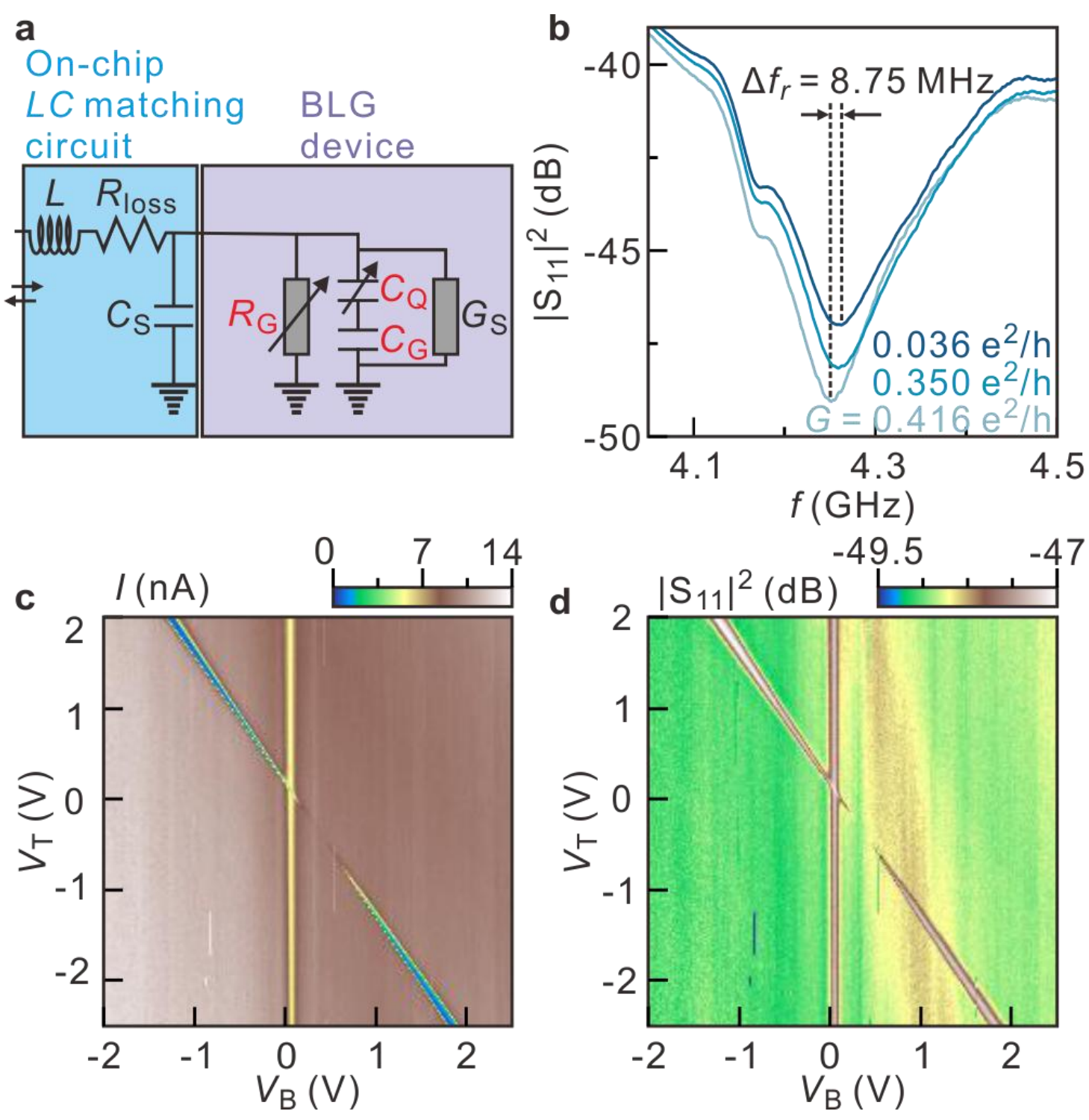

**Figure 2| RF response of the BLG device connected to the superconducting *LC* resonator.** (a) Lumped-element circuit model in which the channel acts as a gate-dependent load on the resonator. (b) RF reflection spectra $|S_{11}|^2$ at three representative gate voltages, ($V_T$, $V_B$) = (2.0 V, -1.41 V), (2.0 V, -1.32 V), and (2.0 V, -1.05 V), with corresponding conductance $G$ = 0.036 (dark blue), 0.350 (medium blue), and 0.416 $e^2/h$ (light blue). The resonance is centered at $f_0 \approx$ 4.251 GHz, and the gate-induced shift between configurations is $\Delta f_r \approx$ 8.75 MHz. (c) DC current map measured as a function of $V_T$ and $V_B$ at $V_{SD}$ = 1 mV. (d) Simultaneously measured RF reflected power $|S_{11}|^2$ at $f_0 \approx$ 4.251 GHz. The RF map reproduces the gate-dependent structure of the DC transport map.

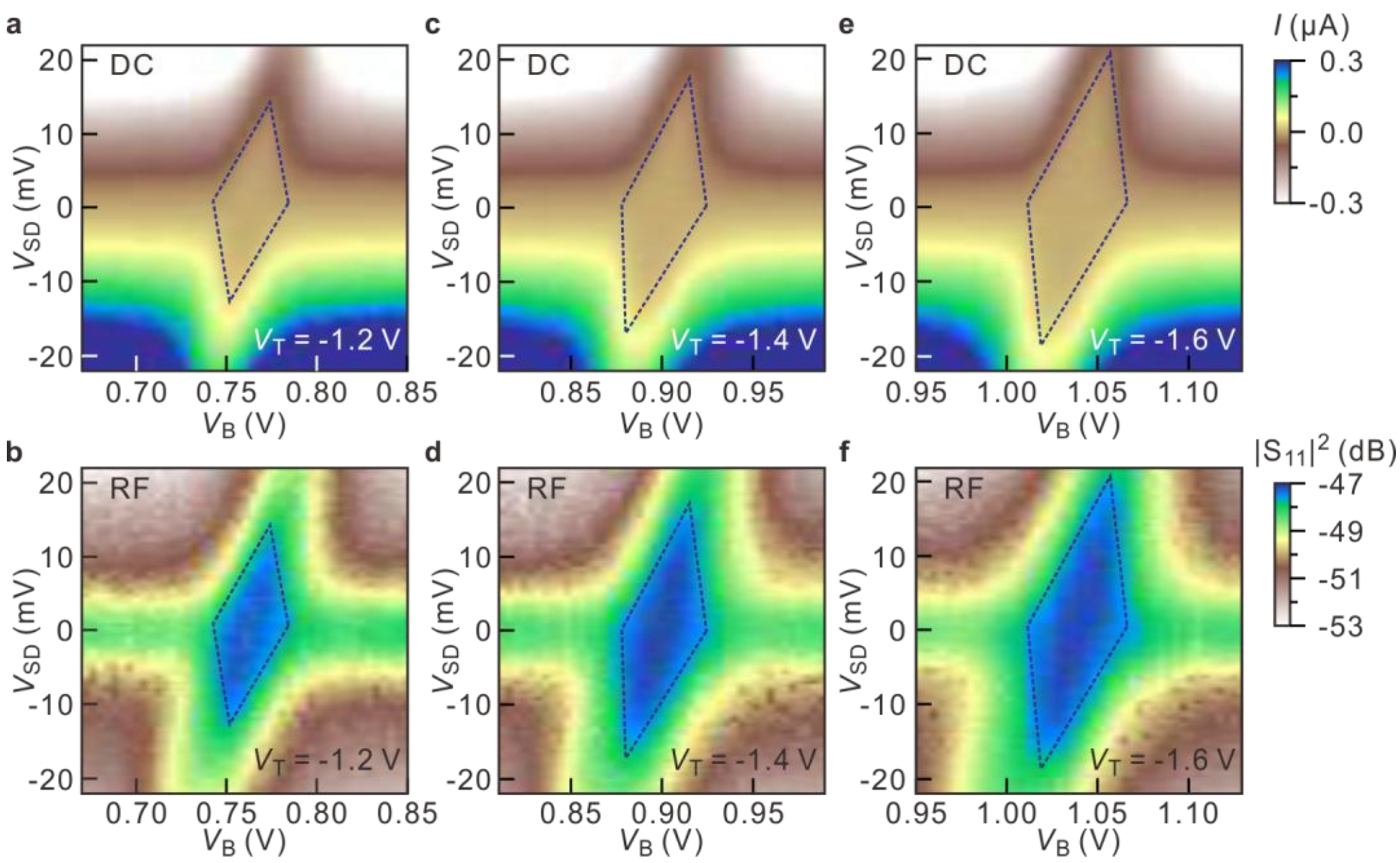

**Figure 3| Finite-bias spectroscopy of the BLG device at *T* = 2 K**. (a), (c), and (e) DC current (*I*) measured as a function of $V_{SD}$ and $V_B$ at $V_T = -1.2$, $-1.4$, and $-1.6$ V, respectively. (b), (d), and (f) Simultaneously measured RF reflected power $|S_{11}|^2$ under the same bias and gate conditions. Navy dashed contours mark the boundary of the suppressed region identified from the finite-bias response. The DC maps display gap-induced regions of suppressed transport, with the RF maps reproducing the corresponding bias- and gate-dependent features.

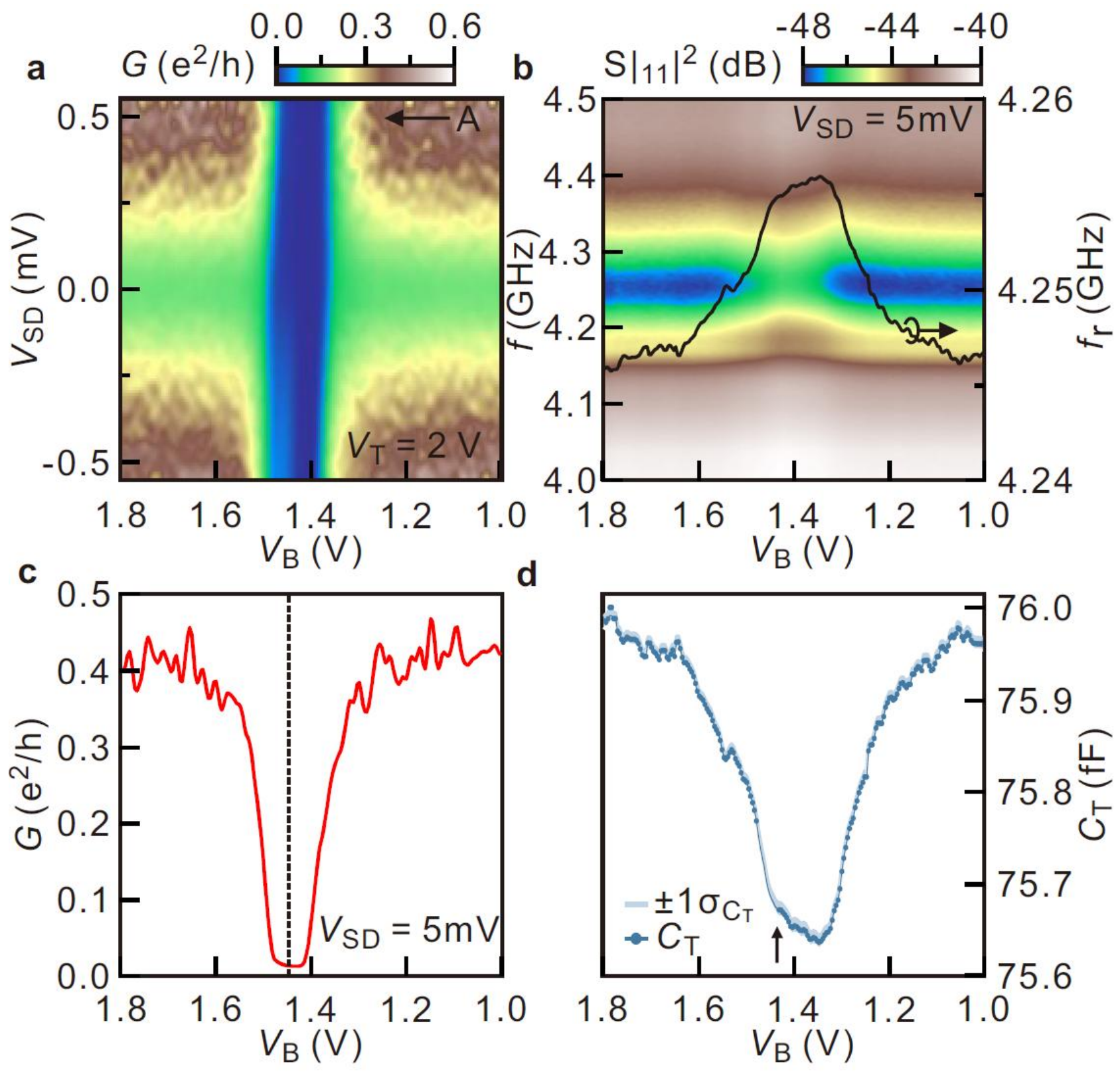

**Figure 4| Capacitance response of BLG near the field-induced gapped regime.** (a) Differential conductance $G$ as a function of $V_B$ and $V_{SD}$ at $V_T = 2$ V. The black arrow marks the line cut A used for the RF measurement. (b) Reflected microwave power $|S_{11}|^2$ along line cut A as a function of frequency and $V_B$ at $V_{SD} = 5$ mV; the black curve traces the extracted resonance frequency $f_r$ (right axis). (c) Line cut of $G$ along the same path at $V_{SD} = 5$ mV. The dashed line marks the conductance minimum. (d) Effective capacitance $C_T$ extracted from the resonance-frequency shift versus $V_B$ (blue markers) with the propagated one-standard-deviation uncertainty ($\pm 1\sigma_{C_T}$, shaded band). The black arrow marks the conductance minimum.

# Supporting Information for "On-chip superconducting GHz RF reflectometry of the capacitance response in bilayer graphene"

Sung Jin An,[1,†] Minseo Cho,[2,†] Minjun Park,[2†] Dohun Kim,[2,†] HyeonJeong An,[1,2] Seung-Bo Shim,[3] Hakseong Kim,[3] Sunghun Lee,[4] Myoung-Jae Lee,[1] Kenji Watanabe,[5] Takashi Taniguchi,[6] Jungpil Seo,[2] Myunglae Jo,[7,*] Youngwook Kim,[2,*] Minkyung Jung[1,8,*]

[1] DGIST Research Institute, DGIST, Daegu 42988, Republic of Korea

[2] Department of Physics and Chemistry, DGIST, Daegu 42988, Republic of Korea

[3] Quantum Technology Institute, Korea Research Institute of Standards and Science, 34113 Daejeon, Republic of Korea

[4] Industrial AX Innovation Research Institute and Sensor Technology Support Center, DGIST, Daegu 42988, Korea

[5] Research Center for Materials Nanoarchitectonics, National Institute for Materials Science, 1-1 Namiki, Tsukuba 305-0044, Japan

[6] Research Center for Electronic and Optical Materials, National Institute for Materials Science, 1-1 Namiki, Tsukuba 305-0044, Japan

[7] Department of Physics, Kyungpook National University, Daegu 41566, Republic of Korea

[8] Department of Interdisciplinary Engineering, DGIST, Daegu 42988, Republic of Korea

[†]These authors contributed equally.

[*]Corresponding authors

Email: myunglae.jo@knu.ac.kr, y.kim@dgist.ac.kr, minkyung.jung@dgist.ac.kr

## Fitting procedure for extraction of the total capacitance

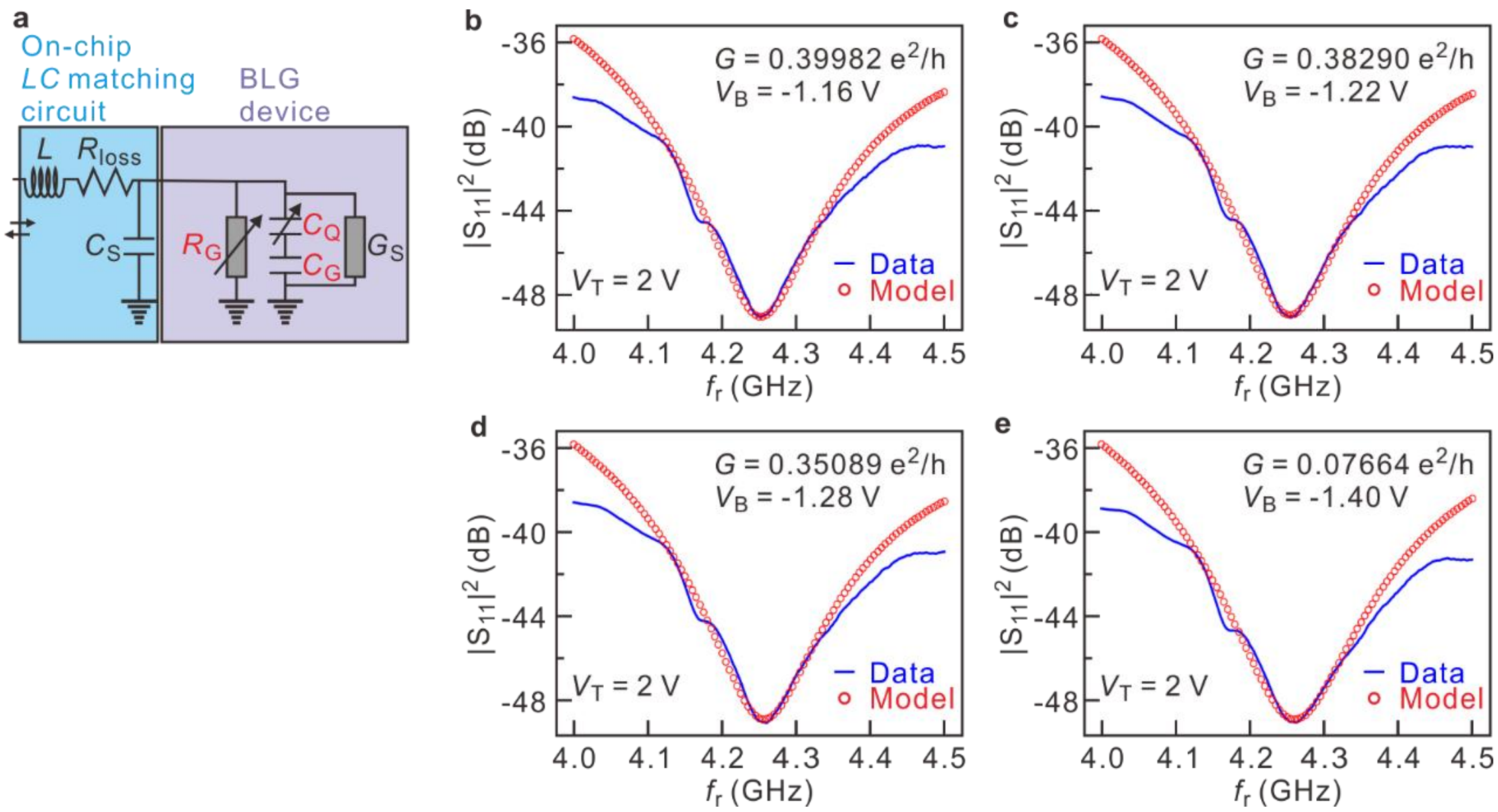

**Figure S1.** (a) Lumped-element circuit model of the device used in Figure 2 of the main text. (b-e) Representative fits of the measured $|S_{11}|^2$ spectra at selected back-gate voltages.

Figure S1(a) shows the equivalent circuit used to analyze the RF-reflectometry data. The BLG device is modeled as a gate-dependent resistance $R_G$ in parallel with the device capacitance $C_t$, where $C_t$ is the series combination of the geometric capacitance $C_G$ and the quantum capacitance $C_Q$. The parasitic shunt capacitance of the on-chip matching circuit, $C_S$, adds in parallel to give the total capacitance $C_T = C_t + C_S$.

We compute the input impedance $Z_{in}$ of this circuit and the reflection coefficient $\Gamma = (Z_{in} - Z_0)/(Z_{in} + Z_0)$, with $Z_0 = 50\ \Omega$, and compare the calculated $|\Gamma|^2$ with the measured $|S_{11}|^2$ spectra to obtain the representative circuit parameters. Dielectric loss in the capacitive branch is represented by an effective loss tangent $\tan\delta_{eff}$, contributing a shunt conductance $G_S = \omega C_T \tan\delta_{eff}$. Representative fits are shown in Fig. S1 (b-e). In the high-conductance regime, the model reproduces the measured spectra well, yielding $L = 18.49$ nH, $\tan\delta_{eff} = 0.0085$, and $R_{loss} = 29.5\ \Omega$, with a total capacitance $C_T$ of approximately 76 fF. Its gate dependence is interpreted in the main text in terms of the quantum-capacitance contribution associated with the BLG density of states.

Near the low-conductance, energy-gap regime, the full line-shape fit becomes less reliable. This deviation likely reflects effects outside the simplified lumped-element model, such as disorder-induced localized states, spatial charge inhomogeneity, and gate-dependent dissipation in the gapped regime. We therefore use the full line-shape fit to fix the representative circuit parameters in the high-conductance regime, and extract the gate-dependent capacitance primarily from the resonance-frequency shift, rather than fitting all circuit parameters independently at every gate voltage. The shaded uncertainty band in Fig. 4(d) was obtained from the standard error of the fitted resonance frequency. The uncertainty in $L$ gives a gate-independent scale uncertainty and is not included in the shaded band.